\documentclass[journal]{IEEEtran}
\ifCLASSINFOpdf
\else
\fi
\usepackage[dvipsnames]{xcolor}
\usepackage[normalem]{ulem}
\usepackage{color,soul}
\usepackage{enumitem}
\usepackage{comment}
\usepackage{longtable}
\usepackage{graphicx}
\usepackage[font=normalsize]{caption}
\usepackage{chemformula}
\usepackage{caption}
\usepackage{booktabs}
\DeclareCaptionLabelSeparator{tableNewline}{\par}
\usepackage{caption}
\usepackage{subcaption}
\usepackage[font={small},skip=10pt]{caption}
\usepackage{booktabs}
\usepackage[english]{babel}
\usepackage{multicol}
\usepackage{fancyhdr}
\usepackage[section]{placeins}
\usepackage{multirow}
\usepackage{flushend}
\usepackage{tabularx}
\usepackage{adjustbox}
\usepackage{float}
\usepackage{lipsum} 
\usepackage{makecell}
\usepackage{booktabs}
\usepackage{array}
\usepackage{hyperref}
\usepackage{amsmath}
\usepackage{amsfonts}
\usepackage{amssymb}
\hypersetup{
  colorlinks=true,
  allcolors=blue,
  allbordercolors={blue!50!black}}
\usepackage[noadjust]{cite}
\usepackage[english]{babel}
\usepackage{graphicx}
\begin{document}
\bstctlcite{IEEEexample:BSTcontrol}
\setlength{\parskip}{0pt}
\title{Molecular Communication-Based Quorum Sensing Disruption for Enhanced Immune Defense}
\author{Shees Zulfiqar and~Ozgur~B.~Akan,~\IEEEmembership{Fellow,~IEEE}
\thanks{The authors are with the Center for neXt-generation Communications (CXC), Department of Electrical and Electronics Engineering, Ko\c{c} University, Istanbul 34450, Turkey (e-mail: \{shali23,akan\}@ku.edu.tr).}
\thanks{O. B. Akan is also with the Internet of Everything (IoE) Group, Electrical Engineering Division, Department of Engineering, University of Cambridge, Cambridge CB3 0FA, UK (email: oba21@cam.ac.uk).}      
\thanks{This work was supported by the AXA Research Fund (AXA Chair for Internet of Everything at Ko\c{c} University).}
\vspace{-1.2mm}
}

\maketitle

\maketitle
\begin{abstract}
Molecular Communication (MC) utilizes chemical molecules to transmit information, introducing innovative strategies for pharmaceutical interventions and enhanced immune system monitoring. This paper explores Molecular communication-based approach to disrupt Quorum Sensing (QS) pathways to bolster immune defenses against antimicrobial-resistant bacteria. Quorum Sensing enables bacteria to coordinate critical behaviors, including virulence and antibiotic resistance, by exchanging chemical signals, known as autoinducers. By interfering with this bacterial communication, we can disrupt the synchronization of activities that promote infection and resistance. The study focuses on RNAIII inhibiting peptide (RIP), which blocks the production of critical transcripts, RNAII and RNAIII, within the Accessory Gene Regulator (AGR) system, thereby weakening bacterial virulence and enhancing host immune responses. The synergistic effects of combining QS inhibitors like RIP with traditional antimicrobial treatments reduce the need for high-dose antibiotics, offering a potential solution to antibiotic resistance. This molecular communication-based approach presents a promising path to improved treatment efficacy and more robust immune responses against bacterial infections by targeting bacterial communication.
\end{abstract}


\section{Introduction}
\label{sec: Introduction}
\IEEEPARstart{M}{olecular} Communication (MC) is the emerging area concerning communication systems where chemical molecules are used to transmit messages \cite{Akan2017FundamentalsOM}. This is a drastic shift away from conventional communication, as MC is organic, a biological phenomenon at the nanoscale. In this system, a transmitter nanomachine emits specific signaling molecules to the surrounding medium, and these molecules spread through the environment up to another receiver nanomachine, which decodes this chemical signal \cite{atakan2012nanoscale}.

For instance, the problem of Antimicrobial Resistance (AMR), which results in bacteria developing mechanisms to counter the effects of antibiotics, is a current worldwide health concern \cite{Aguilar2023Burden}. Conventional antibiotics exert their action on bacterial cells and are commonly ineffective because bacteria develop a resistance mechanism, including enzyme secretion that degrades the antibiotics or forms a biofilm layer as shown in Fig. \ref{1} \cite{soni2024understanding}. Thus, MC-based strategies more attractive, focusing on bacterial QS systems instead of direct bacterial extermination. Interference with QS-related signaling disrupts the bacterial ability to coordinate attack, thus reducing their potency and making them more susceptible to host defenses \cite{cevdet2017molecular}.

    \begin{figure}
        \centering
        \includegraphics[width=1\linewidth]{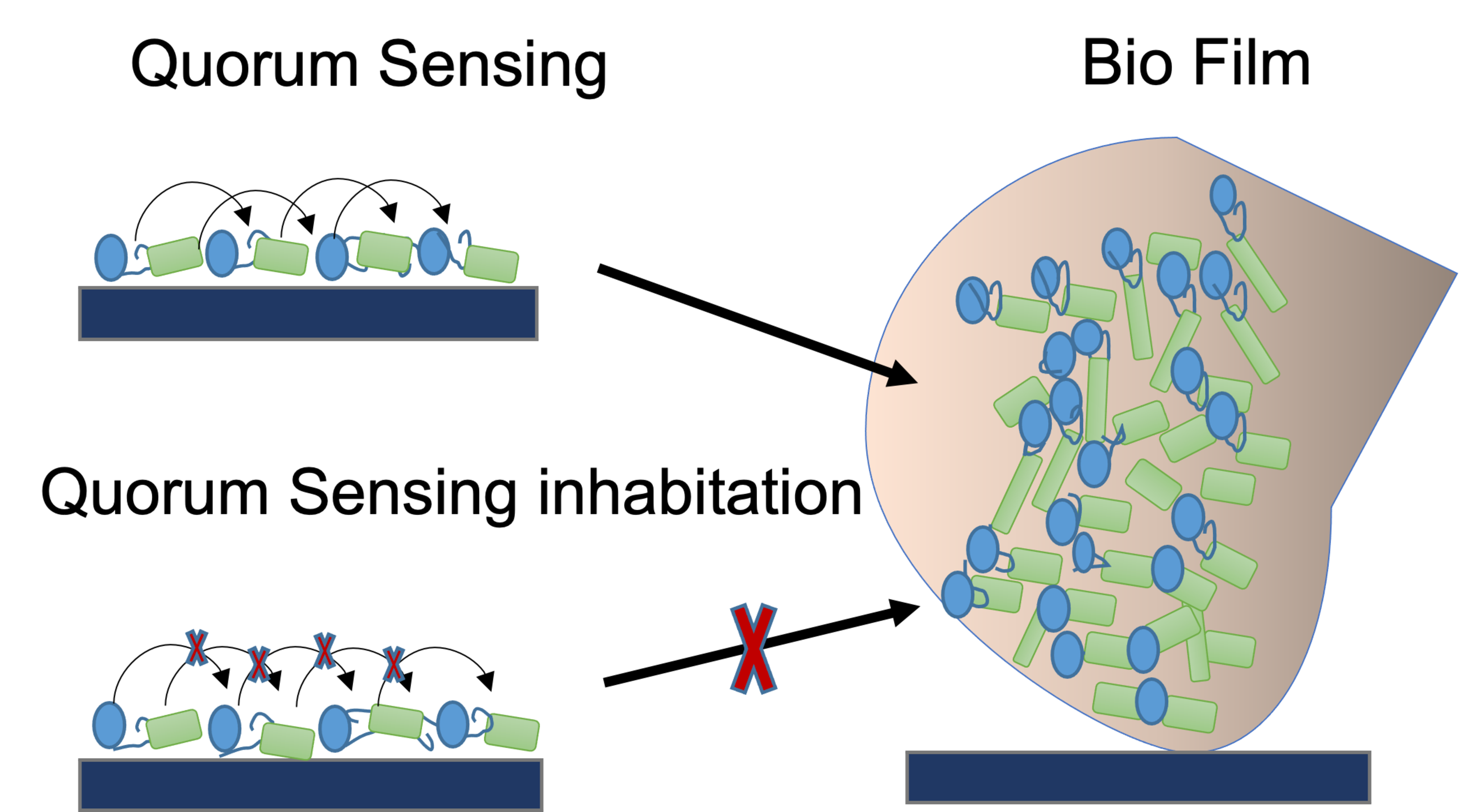}
    \caption{Diagram illustrating the role of quorum sensing in biofilm formation. The top sequence shows quorum sensing, where bacterial communication signals (arrows) lead to biofilm development as cells aggregate. The bottom sequence depicts quorum sensing inhibition, where signal interruption (red crosses) prevents cell communication, inhibiting biofilm formation and maintaining a dispersed bacterial state.}
    \label{fig:1}
\end{figure}
\indent
Quorum Sensing (QS) is particularly relevant within human organisms, where it demonstrates intricate social behaviors via a cell-to-cell signaling method \cite{juszczuk2024molecular}. The adaptive immune system plays a significant role in defending against infections. QS allows bacteria to sense their population density and synchronize behavior changes once a critical cell density is reached \cite{fang2023molecular}. This synchronization affects virulence factor production, biofilm formation, and pathogenic or symbiotic interactions. Bacterial cells can communicate and control the actions of neighboring cells by using chemical substances known as autoinducers \cite{moreno2023quorum}. Thus, when a specific density of these signaling molecules is attained, bacteria turn in phase, and the coordinated actions include forming biofilm and regulating virulence factors. This act of crowd mainly facilitates the pathogenicity of bacteria , more so within areas experiencing high levels of antibiotic resistance for immune system.
\begin{figure}
    \centering
    \includegraphics[width=1\linewidth]{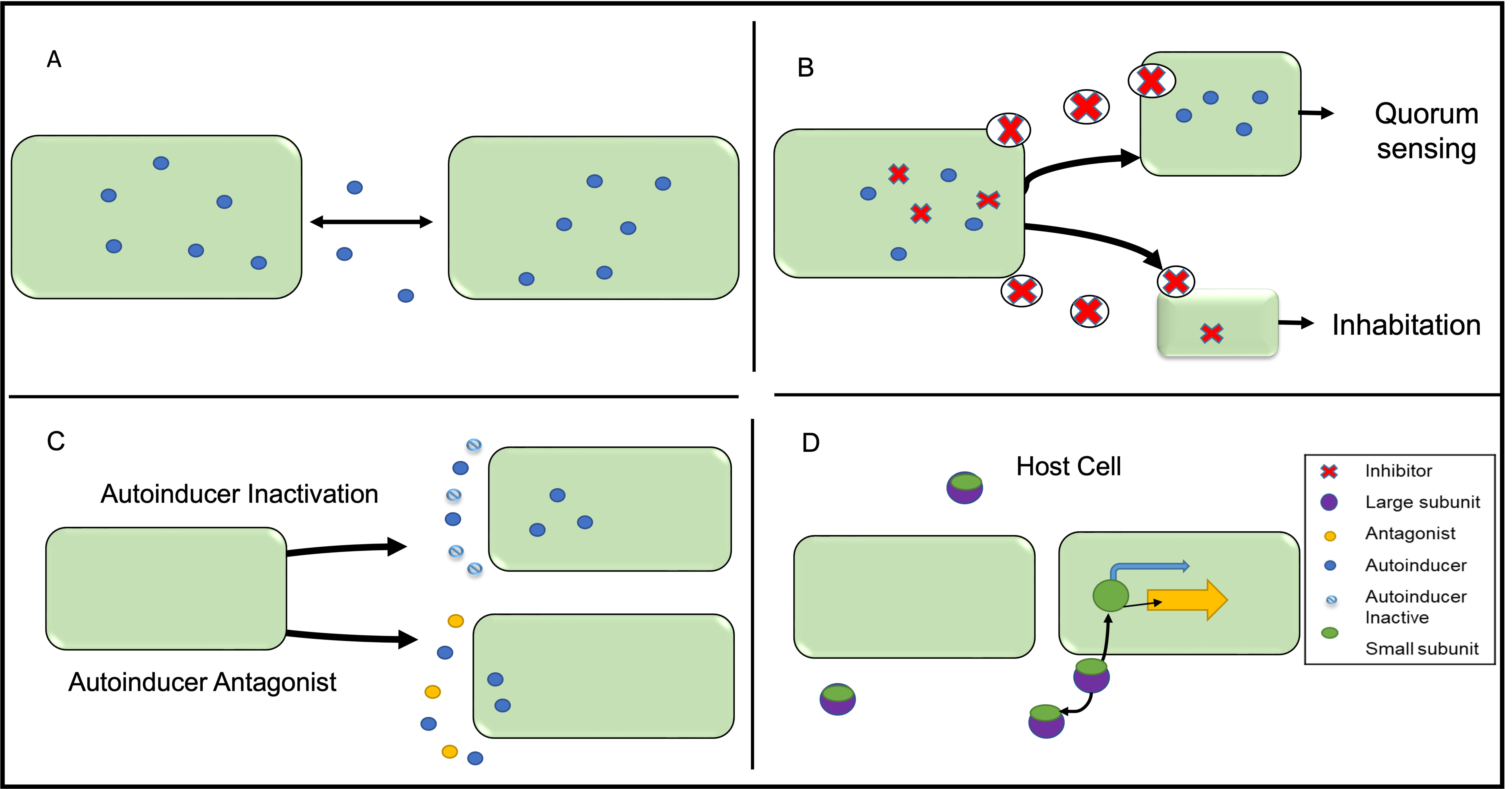}
    \caption{Illustration of cellular communication mechanisms: (A) Cell-cell communication, (B) Quorum sensing and inhibition, (C) Autoinducer inactivation and antagonism, and (D) Host cell interaction with inhibitors.}
    \label{fig:2}
\end{figure}
\indent 

In Gram-positive bacteria, i.e., Methicillin-resistant Staphylococcus aureus (MRSA), peptides serve as signaling molecules known as autoinducers, originating as precursor peptides that are ribosomally produced and converted into active forms via post-translational modifications \cite{lindercharacterization}. An ATP-binding cassette (ABC) transporter mediates the release of these autoinducers, and their accumulation in the environment triggers the communication process by binding to receptors and activating receptor kinases through phosphorylation \cite{kaliniak2024remodeling}. The specific QS systems, such as the AGR system in Staphylococcus species, vary across different Gram-positive bacteria \cite{baer2024single}.

This paper investigate MC-based model that can disrupt QS in bacteria, emphasizing to inhibit quorum sensing in MRSA. This strategy focuses on QS pathways to increase the host's immune system's capacity, decrease dependence on conventional antibiotics, and address the issue of AMR. Section \ref{Transmitter-Receiver model in Quorum sensing} is dedicated to peptide-based mechanisms in the bacteria Staphylococcus. Section \ref{sec:Mathematical for Molecular Communication in Quorum Sensing} presents Mathematical Modeling for bacterial growth and how the immune system responds to toxin production and Section \ref{sec:RESULTS} analyse the behaviour of bacteria over time and how to inactive it. Section \ref{sec:conclusions} conclude the paper.

\section{Transmitter-Receiver model in Quorum sensing}
\label{Transmitter-Receiver model in Quorum sensing}
 Molecular Communication (MC) is based on transmitting information through chemical signals \cite{atakan2012nanoscale}. In bacterial communication, MC is crucial via QS, where bacterial colonies coordinate and control certain general behaviors by producing autoinducers \cite{jana2024quorum}. From an MC perspective, in a bacterial colony, the transmitter is an individual bacterial cell that emits signaling molecules in the environment, and the receiver is the adjacent bacterial cell containing receptors that sense these molecules \cite{singh2024cell}.
In this model, bacteria (acting as transmitters) release autoinducers into the surrounding environment, where these molecules diffuse through the medium. As the bacterial population density increases, the concentration of autoinducers in the environment rises \cite{barbey2024molecular}. Once a threshold concentration is reached, the autoinducers are detected by other bacteria (acting as receivers), triggering specific cellular responses as shown in Fig. \ref{2}. This process enables bacteria to synchronize behaviors across the population, such as biofilm formation, virulence factor production, and antibiotic resistance \cite{das2024electrical}.

The diffusion of these signaling molecules in the environment allows for a scalable and adaptable communication mechanism sensitive to changes in bacterial population density and environmental conditions \cite{koca2024bacterial}. Once detected by receiver bacteria, the autoinducers bind to specific receptors, which may be located on the cell surface or within the cytoplasm \cite{jiang2024bioinspired}. This binding event activates the receptor, leading to downstream signaling cascades that ultimately result in gene transcription and the initiation of coordinated cellular activities \cite{hetta2024quorum}. In Gram-negative bacteria, this often involves interaction with membrane-bound receptors. In contrast, in Gram-positive bacteria, autoinducers are typically transported back into the cell to interact with internal transcription factors, representing a distinct mode of transmitter-receiver communication \cite{atakan2012nanoscale}, \cite{das2024electrical}.
\begin{figure*}[h!]
    \centering
    \includegraphics[width=0.85\linewidth]{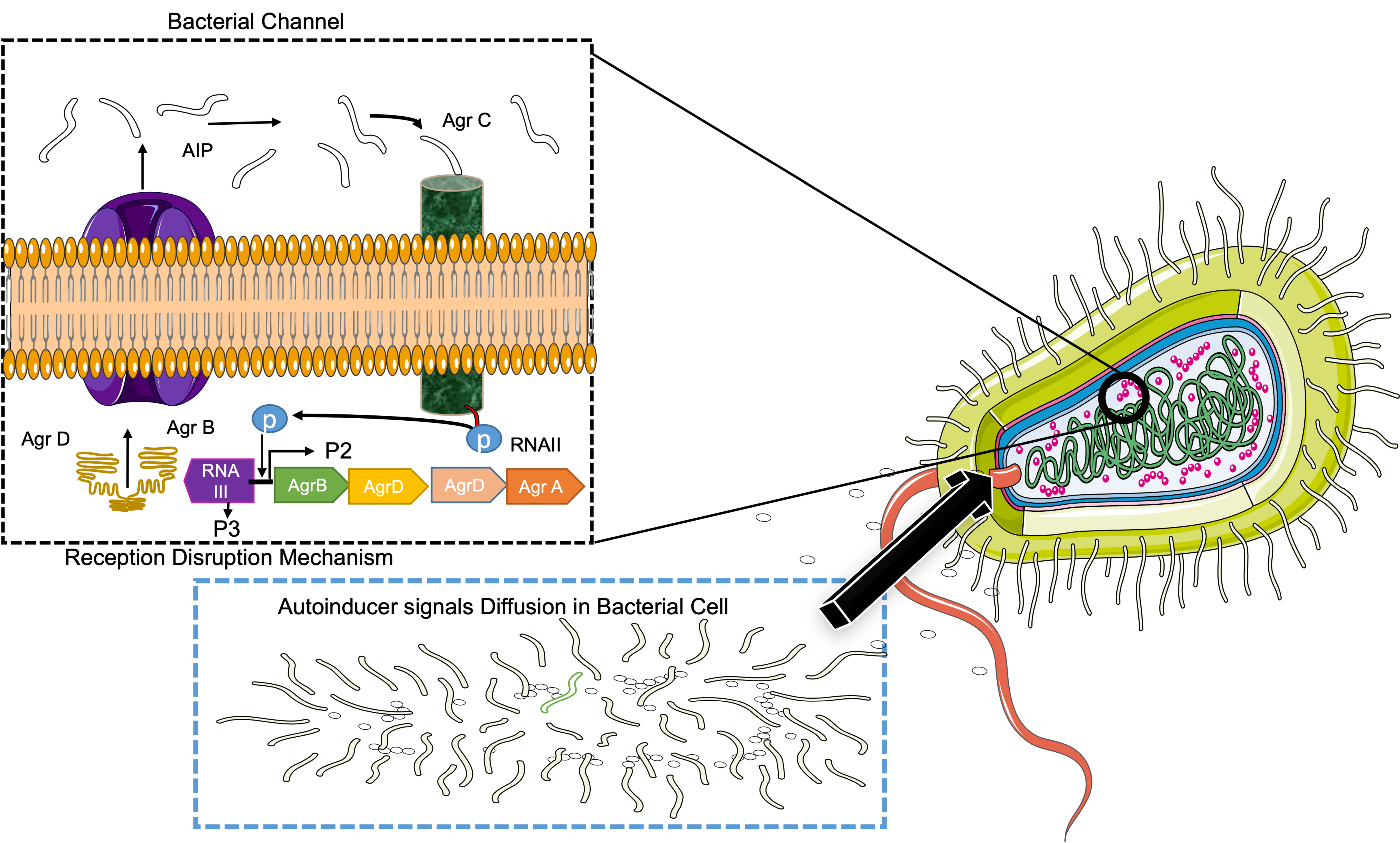}
    \caption{Mechanism of Quorum Sensing (QS) disruption in Staphylococcus aureus through molecular communication. It illustrates the Accessory Gene Regulator (AGR) system, showing the production of autoinducer peptides (AIPs) by AgrD and their secretion via AgrB. AIPs bind to the receptor AgrC, initiating a signaling cascade that activates RNAII and RNAIII.}
    \label{fig:3}
\end{figure*}
\indent 

Methicillin-resistant Staphylococcus aureus (MRSA) represents a group of bacteria that have developed resistance to a wide range of antibiotics, particularly Methicillin and other drugs in the penicillin class \cite{mavdarova2024novel}. This resistance presents a significant challenge for public health due to the limited treatment options. Central to the communication and regulation of bacterial behavior in MRSA is a sophisticated transceiver (Tx-Rx) diffusion-based mechanism \cite{heinlein2024closing}. This system allows bacteria to alter gene expression and behavior in response to changes in cell density through the release and detection of signaling molecules or autoinducers.
\newline
\indent
In Staphylococcus aureus, two critical quorum-sensing systems, SQS 1 and SQS 2, govern the production of toxins. These systems are activated when the concentration of signaling molecules, transmitted and diffused throughout the bacterial environment, reaches a certain threshold as the population grows \cite{moreno2023quorum}. The activation of SQS 1 starts with the transmission of the RNAIII-activating protein (RAP) signal, which initiates toxin production upon diffusion and reception by target molecules. Simultaneously, SQS 2 becomes active during bacterial growth, producing regulatory RNAIII as shown in Fig. \ref{3} \cite{lejars2024bacterial, korem2005transcriptional}. This RNA molecule governs the expression of genes responsible for both cell surface proteins and toxins.

\indent
The synchronization between these two systems ensures bacterial virulence is coordinated with population density. SQS 1, a multisubunit system associated with RAP, operates through a Tx-Rx mechanism \cite{fang2020characterization, tsave2019anatomy}. RAP molecules are transmitted and diffuse through the bacterial environment to interact with TRAP, a sensor molecule embedded in the bacterial membrane. When RAP is sufficiently transmitted and received by TRAP, it triggers TRAP phosphorylation, which is necessary to activate SQS 2. SQS 2 involves a gene cluster known as agr, which becomes active during the mid-exponential growth phase of the bacterium, leading to the production of RNAII and RNAIII  \cite{hetta2024quorum}.

\indent 
RNAIII, a regulatory RNA molecule, plays a key role by encoding toxin genes while repressing cell surface protein genes through RNA-binding proteins \cite{jian2024small}. This diffusion-based communication ensures that the expression of the toxin gene is widespread once a critical population density is reached \cite{rana2024architectures}. Through their Tx-Rx mechanisms, the coordinated action of SQS 1 and SQS 2 establishes a functional signaling network that activates the agr system. This network produces the AIP peptide, which, after sufficient accumulation, diffuses back to interact with the AgrC sensor kinase. This interaction reduces TRAP phosphorylation, releasing TRAP for further interaction with RAP, which, in turn, triggers the AGR system to synthesize toxins, demonstrating the synchronization of SQS 1 and SQS 2 during the early stages of quorum sensing regulation \cite{vadakkan2024review}.

\indent 
The external signaling system in Staphylococcus aureus, known as the Agr quorum sensing system, is a highly organized Tx-Rx diffusion-based communication network. This system regulates genetic behavior critical for bacterial virulence. AgrD, the pro-peptide of AIP, is processed and transmitted by AgrB to produce AIP. Once secreted into the environment and accumulating to a specific concentration, AIP diffuses back and rebinds to the AgrC sensor kinase, causing its autophosphorylation   \cite{fang2023molecular}. AgrA receives this signal, the response regulator, which is then phosphorylated by AgrC, initiating the transcription of target genes like RNAIII, the primary effector molecule of the agr system. RNAIII controls the expression of several other genes involved in virulence factors and other regulatory pathways.

Promoters P2 and P3, activated by the phosphorylated AgrA, drive the transcription of the agr operon (agrBDCA) and RNAIII  \cite{yamaguchi2024physalin}. Additionally, the SrrA/SrrB and SaeR/SaeS two-component systems modulate the activity of the Agr system in response to environmental stresses detected through Tx-Rx communication. Sara, a global regulator, also interacts with the Agr system, modulating virulence gene expression. At the same time, metabolic signals from CcpA and CodY, received through the Tx-Rx system, influence the Agr system based on nutrient availability. Moreover, phenol-soluble modulins (PSMs), short peptides involved in S. aureus pathogenicity, are controlled by the Agr system via this diffusion-based communication network \cite{morganbryan}.
\newline
\indent
Understanding the intricacies of these Tx-Rx diffusion-based quorum sensing systems is crucial for developing new antibiotics to combat MRSA infections. The complex interplay of signaling molecules and their coordinated response in bacterial populations highlights the sophisticated nature of bacterial communication and virulence regulation.

\section{Mathematical for Molecular Communication in Quorum Sensing}
\label{sec:Mathematical for Molecular Communication in Quorum Sensing}
Bacterial communication, known as Quorum Sensing (QS), depends on cell density and triggers collective bacterial behaviors when a critical population threshold is reached \cite{boedicker2009microfluidic}. In the diffusion-based transceiver (Tx-Rx) model of QS, bacterial populations act as both transmitters and receivers, exchanging information through the secretion and diffusion of autoinducers—small signaling molecules essential to QS \cite{martins2019computational}.

\subsection{Diffusion of Signaling molecules}
\label{subsec: Diffusion of Autoinducers}

The pathways of signaling molecules such as RAP (Regulator of Activator Protein) and AIP (Autoinducer Peptide) in bacterial environment can be analyzed using Fick’s second law of diffusion where the rate of change of concentration with time depends upon the diffusion gradient \cite{akyildiz2019information},

\begin{equation}
\frac{\partial C(r,t)}{\partial t} = D \nabla^2 C(r,t) + S(r,t),
\label{1}
\end{equation}
where $C(r,t)$ is the concentration of signaling molecule (AIP) at position $r$ and time $t$, $D$ is the diffusion coefficient, $\nabla^2 C(r,t)$ describes spatial diffusion, and $S(r,t)$ represents the autoinducer production rate by bacteria \cite{efendiev2013existence}.

In biological environments, $D$ varies with position $r$, reflecting differences in medium properties like viscosity. This variability is modeled as

\begin{equation}
\frac{\partial C(r,t)}{\partial t} = \nabla \cdot \textbf{(}D(r) \nabla C(r,t)\textbf{)} + S(r,t),
\label{2}
\end{equation}

At cell membranes, autoinducers may undergo reactions or degradation captured by the reactive boundary condition.

\begin{equation}
-D(r) \frac{\partial C(r,t)}{\partial n} \bigg|_{\Sigma} = k_r C(r,t)\bigg|_{\Sigma},
\label{3}
\end{equation}
where $\frac{\partial C(r,t)}{\partial n}$ is the concentration gradient average to boundary $\Sigma$, and $k_r$ is the reaction rate constant.

Additionally, the production of autoinducers is often time-dependent \cite{fu2024deciphering}, modeled as:

\begin{equation}
S(r,t) = S_0(r) f(t),
\label{4}
\end{equation}

$S_0(r)$ is the spatial source distribution, and $f(t)$ models temporal variations due to environmental factors or bacterial growth cycles.

Autoinducers can also be transported via bulk fluid flow (convective transport) \cite{benbelkacem2024should}, integrated into the model as,

\begin{equation}
\frac{\partial C(r,t)}{\partial t} + \mathbf{v}(r) \cdot \nabla C(r,t) = \nabla \cdot (D(r) \nabla C(r,t)) + S_0(r) f(t),
\label{5}
\end{equation}
where $\mathbf{v}(r)$ is the fluid's velocity field carrying the autoinducers. This combination of diffusion and convection provides a comprehensive view of autoinducer distribution.

\subsection{Signal Reception and Activation of QS}
 In the SQS 1 mechanism, signaling molecule RAP (Regulator of Activator Protein) binds to its specific receptor TRAP (Transcriptional Regulator of the Agr System) \cite{shaw2007inactivation, gov2001rnaiii}. This interaction can be best described through a finite mathematical model that gives the dynamic equilibrium of phosphorylated and unphosphorylated TRAP. In this system, specific RAP molecules can interact with TRAP by attaching to its surface such that after phosphorylation of the TRAP, it hides the TRAP but simultaneously exposes unphosphorylated TRAP.

Mathematically, we define the expected concentration of RAP at position \( x \) at time \( t \) is represented by \( C(x,t) \). \( T \) is the concentration of unphosphorylated TRAP, and \( T_p \) denotes the accumulation of phosphorylated TRAP. The dynamics of this interaction involve two key processes: the activation of TRAP by RAP and the later neutralization of the activated TRAP by removing the phosphate group from the product to reform the initial inactive TRAP.

 The formation process can be analyzed mathematically in terms of formation rate $k_1 C(x,t) T$ , \( k_1 \) is the rate constant for the phosphorylation reaction. On the other hand, dephosphorylation decreases the levels of the phosphorylated species of TRAP and is described as $k_2 T_p $, \( k_2 \) is the rate constant for the dephosphorylation process.

When dividing these two processes, the specific differential equation corresponding to the concentration of phosphorylated TRAP and change in time is obtained. The rate of change of phosphorylated TRAP is given by the difference between the phosphorylation and dephosphorylation rates, leading us to the equation:
\begin{equation}
\frac{d T_p}{dt} = k_1 C(x,t) T - k_2 T_p,
\label{6}
\end{equation}

In the SQS 2 mechanism, a substantial engagement of AIPs (Autoinducer Peptides) with the AgrC sensor is involved in bacterial quorum sensing. The more AIP molecules collect in the environment; they form a complex with the receptor AgrC that initiates a set of biochemical changes that aim at activating the Agr system. The dynamics of this interaction can be modeled using the following differential equation:
\begin{equation}
\frac{d A_p}{dt} = k_3 C(x,t) A - k_4 A_p,
\label{7}
\end{equation}
where \( A_p \) and \( A \) portray phosphorylated and non-phosphorylated AgrC concentrations, respectively. The parameter \( k_3 \) is the rate constant for binding AIP to AgrC and phosphorylating AgrC, and it is an index of how efficiently AIP molecules foster this event. On the other hand, \( k_4 \) represents the rate constant of dephosphorylation of AgrC, which is the process where phosphorylate AgrC returns to its basal non-phosphorylated state.

A higher density of an AIP favors binding interactions with AgrC, resulting in increased phosphorylation of AgrC. At the same time, the presence of \( k_4 A_p \) suggests that the level of phosphorylated AgrC reduces due to the action of the phosphatase enzyme. Therefore, the reciprocal phosphorylation between AIP and AgrC demonstrates a dynamic equilibrium between the process of phosphorylation and dephosphorylation, which is critical to the modulation of the bacterial signaling pathways related to the comprehension of algae communication and activation of defense mechanisms in bacteria.
\subsection{Gene Regulation via RNAIII}
The general regulatory protein RNAIII regulates the expression of virulence genes in the AGR system. The following differential equation can describe the production of RNAIII:

\begin{equation}
\frac{dR}{dt} = k_5 A_p - k_6 R,
\label{8}
\end{equation}
where \( R \) represents the concentration of RNAIII, \( k_5 \) is the rate constant for RNAIII transcription activated by phosphorylated AgrA, and \( k_6 \) is the degradation rate of RNAIII. This model suggests that the concentration of RNAIII increases with the level of phosphorylated AgrA while simultaneously decreasing due to its degradation.

The coupling between SQS 1 and SQS 2 ensures a synchronized response to population density. This feedback mechanism can be represented by the following equations:

\begin{equation}
    \frac{dT_p}{dt} = k_1 C(x,t) T - k_2 T_p + f(R),
    \label{9}
\end{equation}

\begin{equation}
\frac{dA_p}{dt} = k_3 C(x,t) A - k_4 A_p + g(T_p),
\label{10}
\end{equation}
where \( f(R) \) and \( g(T_p) \) feedback terms modulate the dynamics of TRAP and AgrC by the levels of RNAIII and phosphorylated TRAP, respectively. This loop underscores the complex mechanisms of controlling the signaling pathways and action regarding the number of concentrations of molecules involved.

 Steady-state RNAIII levels required to activate the virulence genes. This activation can be modeled using a Hill function to represent the cooperative binding of RNAIII in regulating gene expression:

\begin{equation}
V(R) = \frac{R^n}{K^n + R^n},
\label{11}
\end{equation}
where V(R) denotes the expression level of virulence genes; R, RNAIII concentration; K, the concentration of RNAIII required to elicit half maximum response; and n, Hill coefficient, which determines the extent of cooperativity among RNAIII molecules in gene expression regulation. \indent 

RNAIII concentration controls virulence genes whose dysregulation impairs the immune system's ability to identify the presence and infection of bacteria. Suppose that the synthesis and accumulation of RNAIII become more significant than the threshold value. In that case, the pathogen may release higher factors to promote its virulence, overcome the host immune system, and cause more severe disease. This increased virulence not only suppresses immune conditions, but also causes long-term infections and higher pathogenicities of the disease.

\section{Results and Discussions}
\label{sec:RESULTS}
This section shows how bacteria control processes concerning density and the environment within this density. It learns to produce and detect signaling molecules called autoinducers; it provides knowledge of gene regulation that results in various coordinated behaviors when a specific concentration of autoinducers is reached.
\indent

\begin{figure}[h!]
    \centering
        \includegraphics[width=1\linewidth]{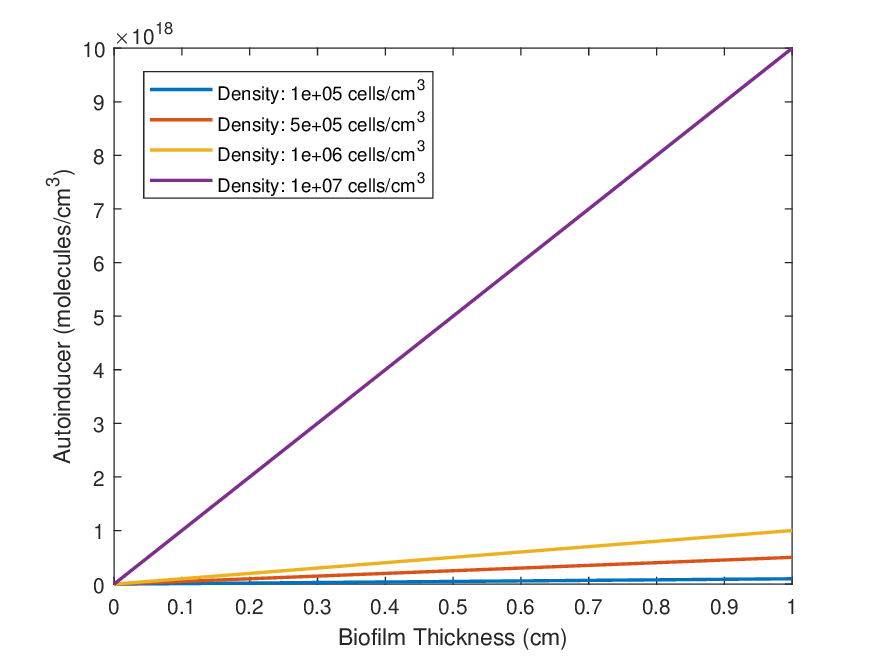}
    \caption{Relationship between autoinducer concentration and biofilm thickness in a diffusion model with varying cell densities.}
    \label{fig:5}
\end{figure}

In Sect. \ref{Transmitter-Receiver model in Quorum sensing}, molecular communication mechanism that considers diffusion of autoinducer to pass the threshold is explained while the simulation result depicts in Fig. \ref{fig:5} demonstrate that cells of MRSA at a high concentration such as 500 cells/ml increase the concentration of autoinducer to the quorum sensing limit of 1$\times 10^6$ molecules/cm\textsuperscript{3}, reaching the quorum sensing threshold (1$\times 10^6$ molecules/cm\textsuperscript{3}) within approximately 9–10 hours, which enables the bacteria to coordinate virulence responses and potentially evade the immune system. However, at dilute cell concentrations (10, 50, and 100 cells/mL), the autoinducer concentration does not exceed this threshold during 24 hours, thus inhibiting the MRSA gear behaviors, such as biofilm formation. 
\begin{figure}[h!]
    \centering
        \includegraphics[width=1\linewidth]{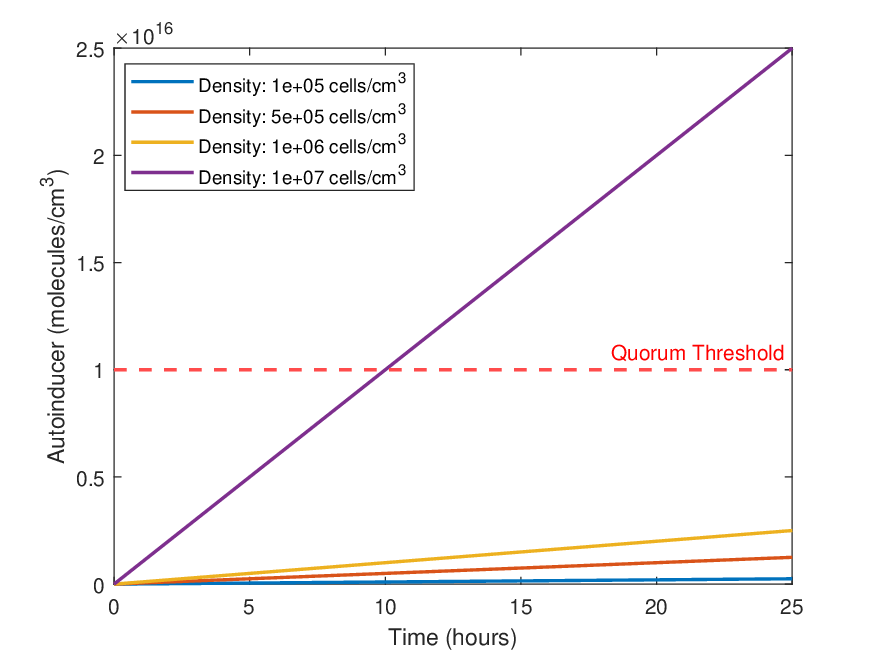}
    \caption{Role of Production Rate and Cell Concentration on Quorum Sensing in MRSA. The plot shows the autoinducer concentration over time at various cell densities, highlighting when the quorum sensing threshold (1$\times 10^6$ molecules/cm\textsuperscript{3}) is reached}
    \label{fig:4}
\end{figure}
\begin{table}[h!]
\centering
\caption{Autoinducer Concentration as a Function of Cell Density and Biofilm Thickness for MRSA}
\label{table:autoinducer_concentration}
\begin{tabular}{|> {\centering \arraybackslash} m{1.3cm}|> {\centering \arraybackslash} m{2.1cm}|> {\centering \arraybackslash} m{1.3cm}|}
\hline
\textbf{Density}&\textbf{BiofilmThickness}&\textbf{Autoinducer}\\ \hline
10             & 0.2                 & $1.0 \times 10^5$    \\ \hline
10          & 0.5           & $2.5 \times 10^5$       \\ \hline
50        & 0.2         & $5.0 \times 10^5$  \\ \hline
50       & 0.5                     & $1.25 \times 10^6$     \\ \hline
100      & 0.2                & $1.0 \times 10^6$   \\ \hline
100       & 0.5           & $2.5 \times 10^6$   \\ \hline
500    & 0.2         & $5.0 \times 10^6$        \\ \hline
500         & 0.5    & $1.25 \times 10^7$        \\ \hline
500      & 1.0         & $2.5 \times 10^7$            \\ \hline
\end{tabular}
\end{table}
Fig. \ref{4} demonstrates that in MRSA (Methicillin-Resistant \textit{Staphylococcus aureus}), the autoinducer concentration increases with biofilm thickness and is particularly elevated at higher cell densities, such as 500 cells/ml. This rise in autoinducer concentration enables MRSA to reach the quorum sensing threshold more rapidly, which triggers collective behaviors like biofilm formation, toxin release, and immune evasion. In high-density biofilms, MRSA coordinates these defenses effectively, making it challenging for the immune system to penetrate and eliminate the bacteria. In contrast, at lower cell densities (e.g., 10, 50, or 100 cells/ml), the autoinducer concentration remains below the quorum sensing threshold across practical biofilm thicknesses, limiting MRSA’s ability to coordinate responses and making it more vulnerable to immune attack. This relationship between cell density, biofilm thickness, and quorum sensing activation highlights potential therapeutic strategies, such as quorum sensing inhibitors, which could prevent MRSA from achieving the critical autoinducer concentration necessary for effective communication and coordinated defense, thereby enhancing the immune system's ability to control and eliminate the infection.

We examine the diffusion behavior of autoinducing peptides (AIPs) in \textit{Staphylococcus aureus} and its relation to quorum sensing and immune evasion. We perform simulations to compare how different diffusion coefficient ($D$), production rate ($S_0$), and threshold concentration ($C_\text{th}$) values affect AIP concentration profiles with distance to a bacterial colony. 
\begin{figure}[h!]
    \centering        \includegraphics[width=1\linewidth]{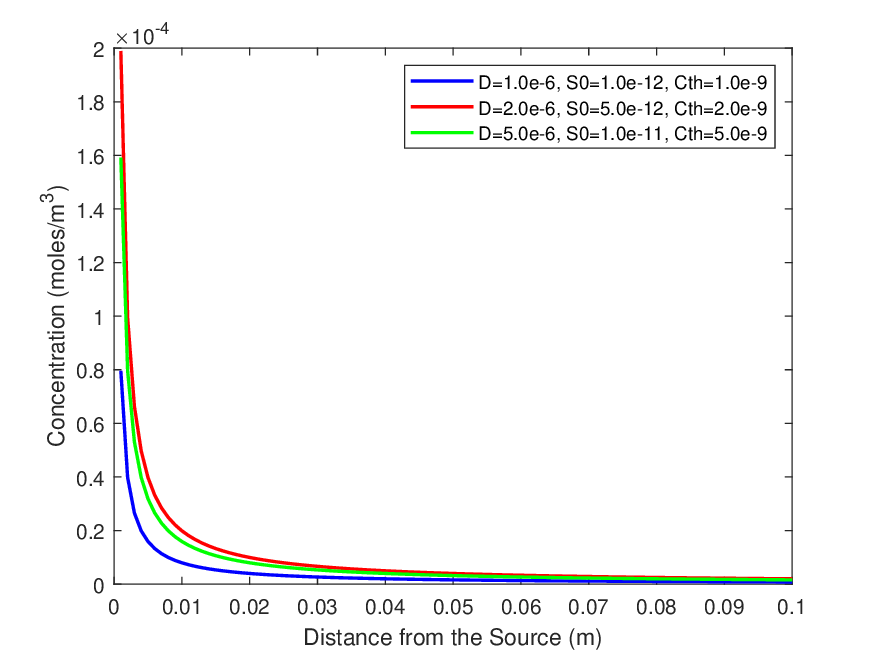}
    \caption{Concentration profiles of autoinducing peptides (AIPs) as a function of distance from the source in a diffusion model with varying parameters. The curves represent different conditions: $D=1.0 \times 10^{-6} \, \text{m}^2/\text{s}$, $S_0=1.0 \times 10^{-12} \, \text{mol/s}$, $C_\text{th}=1.0 \times 10^{-9} \, \text{mol/m}^3$ (blue); $D=2.0 \times 10^{-6} \, \text{m}^2/\text{s}$, $S_0=5.0 \times 10^{-12} \, \text{mol/s}$, $C_\text{th}=2.0 \times 10^{-9} \, \text{mol/m}^3$ (red); and $D=5.0 \times 10^{-6} \, \text{m}^2/\text{s}$, $S_0=1.0 \times 10^{-11} \, \text{mol/s}$, $C_\text{th}=5.0 \times 10^{-9} \, \text{mol/m}^3$ (yellow). }
    \label{fig:6}
\end{figure}

\begin{table}[h!]
\centering
\caption{Parameter values for diffusion coefficient and production rate used in the diffusion model for autoinducing peptides (AIPs) in \textit{Staphylococcus aureus}.}
\begin{tabular}{|c|c|}
\hline
\textbf{Diffusion Coefficient} ($D$) & \textbf{Production Rate} ($S_0$) \\ \hline
$1.0 \times 10^{-6} \, \text{m}^2/\text{s}$ & $1.0 \times 10^{-12} \, \text{mol/s}$ \\ \hline
$2.0 \times 10^{-6} \, \text{m}^2/\text{s}$ & $5.0 \times 10^{-12} \, \text{mol/s}$ \\ \hline
$5.0 \times 10^{-6} \, \text{m}^2/\text{s}$ & $1.0 \times 10^{-11} \, \text{mol/s}$ \\ \hline
\end{tabular}
\end{table}

Our findings also shows that when diffusion coefficients increased, the concentration of AIP decreased with distance at a higher rate, meaning that a higher diffusion rate may impair quorum sensing activation unless the bacteria manufacture more AIP to compensate for this rate. The concentration at which quorum sensing activation is achieved, $C_\text{th}$, is reached faster with higher production rates; hence, quorum sensing occurs farther from the source. This implies that to maintain efficiency in quorum sensing in the high-diffusion environment, \textit{S. aureus} produces relatively higher levels of AIP.

As soon as the agr system is activated, \textit{S. aureus} enhances the secretion of virulence factors that allow it to evade the immune system. These include proteins that avoid immune recognition and the formation of biofilms that shield the colony from immune system attack. 
Fig. \ref{7} represents the Staphylococcus aureus signaling and agr activity, which is affected by the concentration of AIP (1.0e-9M, 1.0e-8 M, and 1.0e-7 M). At 1.0e-9 M, the pre-AIP activity remains negligible, and only a slightly elevated activity is found during exposure, pointing to weak signal interaction. Yet, at 1.0e-8 M and, primarily at 1.0e-7 M, agr activity increases during the exposure to AIP, with higher cell-to-cell signaling observed. Internalized expression also increases activity since it triggers the cells to produce virulence factors that encourage necessary group behavior in infection.
In post-AIP, a similar yet more prolonged increase in agr activity correlates with higher concentrations (especially 1.0e-7 M), indicating that even without the AIP signal, cells continue to communicate. This sustained response assists S. aureus in staying active, preparing the biofilm and the immune-evasive proteins to counter the host immune response and make the infections more resistant to the host immunity.
\begin{figure}[h!]
    \centering
    \includegraphics[width=1 \linewidth]{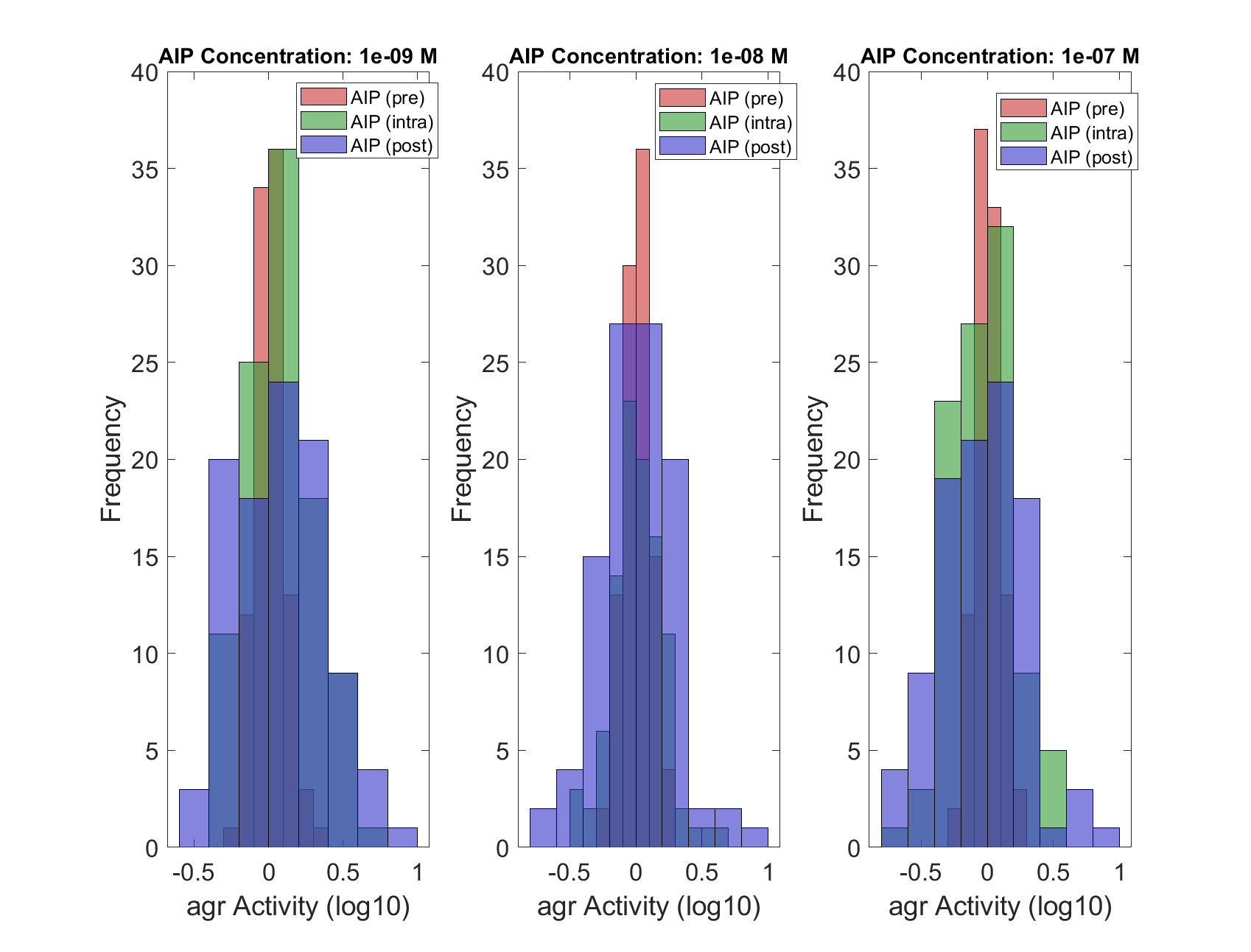} 
    \caption{Distribution of \textit{agr} activity in \textit{Staphylococcus aureus} at varying AIP (autoinducing peptide) concentrations. The three panels represent different AIP concentrations: $1.0 \times 10^{-9} \, \text{M}$ (left), $1.0 \times 10^{-8} \, \text{M}$ (center), and $1.0 \times 10^{-7} \, \text{M}$ (right).}
    \label{fig:7}
\end{figure}

\begin{figure} [h!]
    \centering
        \includegraphics[width=1\linewidth]{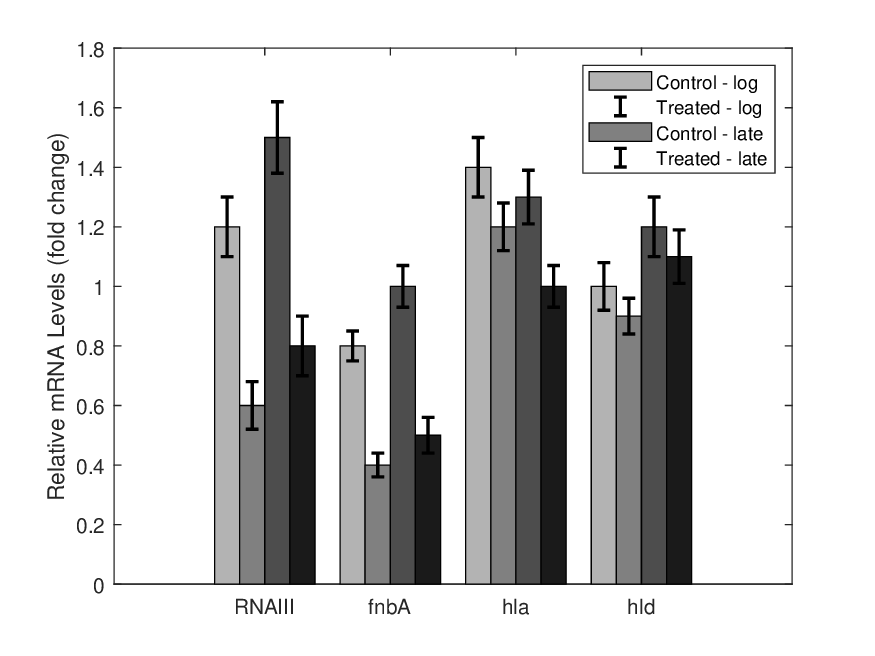}
    \caption{Relatives expression of genes (RNAIII, fnbA, hla, hld) of Staphylococcus aureus under various conditions.}
    \label{fig:8}
\end{figure}

\begin{figure} [h!]
    \centering
        \includegraphics[width=1\linewidth]{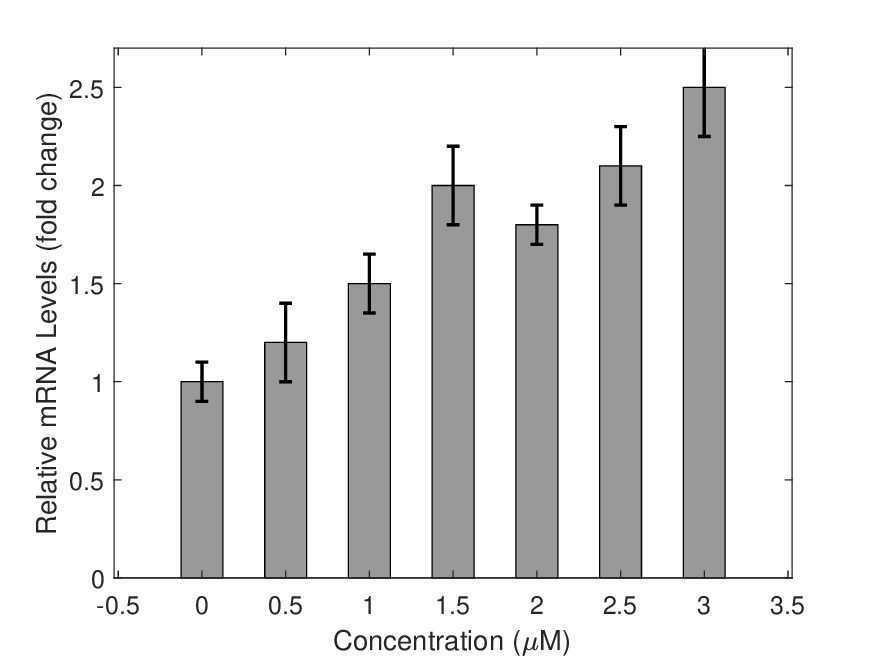}
    \caption{A comparison of the quantity of a specific gene in Staphylococcus aureus at the mRNA level at different concentrations (0 to 3 µM).}
    \label{fig:9}
\end{figure}
 Fig. \ref{8} shows that the number of mRNAs differed between control and treated samples in both log and late growth phases. This indicates that the treatment supports the down regulation of key markers such as FNbA and RNAIII, which are involved in bacterial interaction and pathogenicity. However, hla  declines moderately, suggesting that the response to treatment is complex. However, after treatment, overexpression of the hld gene may cause increased bacterial interaction, hence increasing pathogenicity and biofilm formation.
 
\indent
Fig. \ref{9} shows the direct correlation between the changes in mRNA levels and the concentration; however, there is a small inhibition of 1.5µM. The line joining error bars depict standard deviation to emphasize different degrees of measurement. This implies that intermediate concentrations enhance the expression of genes that can help regulate microbe signals and group conduct.
 
Fig. \ref{10} depicts that with improved concentration of the treatment there is better biofilm inhibition and reduced cell growth. At such levels of concentration, the inhibition of biofilm is relatively small while cell growth rate is great. Nevertheless, as the concentration is increased, biofilm inhibition is much greater averaging 80\% with cell growth reduced to about 30\%. This suggests that the treatment can interfere with bacterial biofilms reducing the likelihood that bacteria will be able to avoid an immune response as often happens if bacteria form biofilms. 
\begin{figure}[h!]
           \includegraphics[width=1\linewidth]{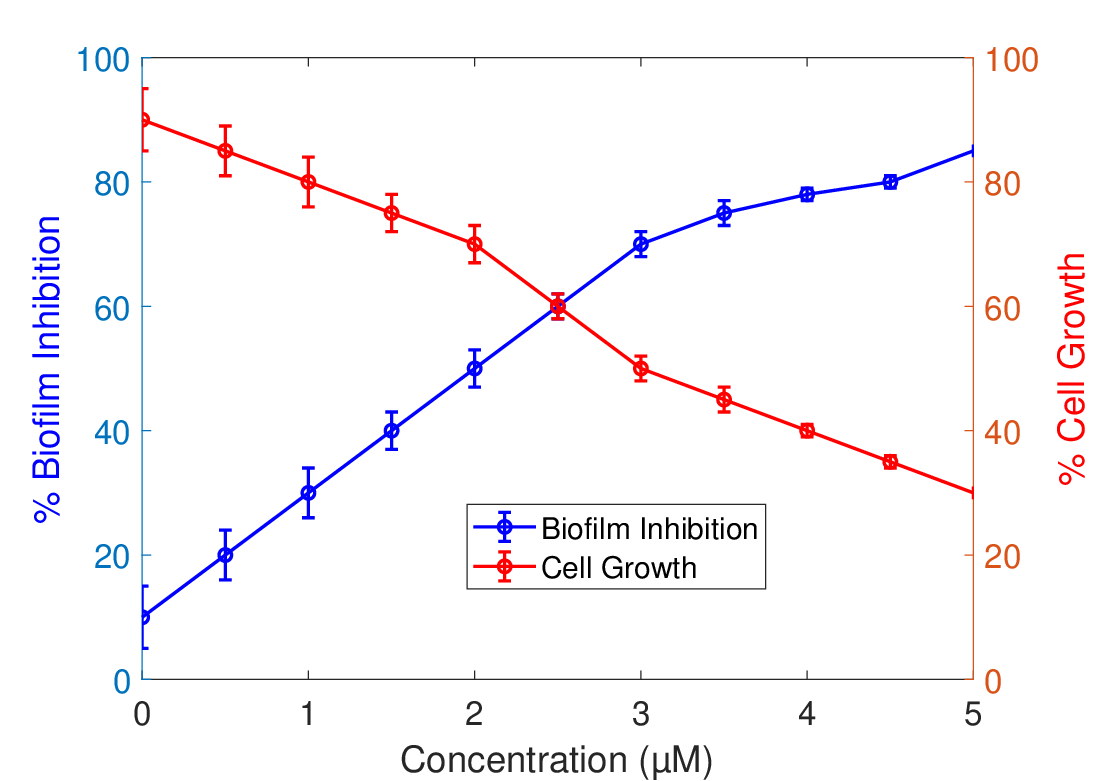}
        \caption{Relationship between treatment concentration and its effects on biofilm inhibition and cell growth.
 }
        \label{fig:10}
    \end{figure}
    \indent 
    \newline 
    \indent 
    Since it reduces bacterial growth and strengthens biofilm structure, the treatment could improve immune- mediated clearance but the high concentrations of the agent should be evaluated for their impact on host cells.Such information is highly valuable when it comes to defining the relationship between concentration levels and gene regulation in bacteria pathogenicity hence the possible therapeutic approaches to infected diseases. By extension, it could be suggested that ratios might be selectively adjusted to ensure the best results are achieved in clinical practice.

\section{conclusions}
\label{sec:conclusions}
Molecular communication  has been identified as a potential approach for nano-scale communication and closely aligns with nanotechnology and precision medicine applications. QS within bacterial populations is a paradigm where bacteria interact and regulate functions such as biofilm formation and pathogenicity using QS molecules known as autoinducers. Illustrates the necessity of quorum sensing in Staphylococcus aureus infections and its relationship to immune resistance and staphylococcal growth control. Using the data obtained from the simulations, we further described how different concentrations of AIPs impact the activation of the agr system in S. aureus and, consequently, virulence factors. Differences in AIP concentrations also make the agr gene respond with higher strength and longer duration to increase the virulence and achieve the entire bacterial colony behavior to escape from the immune system, as depicted in the bacterial growth curves.
Conversely, low AIP levels mean less agr activation, subsequently leading to efficient checks and balances by the immune system on bacterial proliferation. This means that utilizing quorum sensing density can easily explain the level at which bacterial growth is encouraged and the immune system subdued. The blockade of RNAIII or the interruption of AIP signaling has been described as a promising approach to attenuate agr activation, shake off S. aureus virulence, and promote immune efflux. The knowledge of quorum sensing and immune interactions adds to the framework of likely anti-bacterial strategies that interrupt bacterial communication.

\bibliography{references}
\bibliographystyle{ieeetr}
\end{document}